\documentclass[aps,prb,twocolumn,showpacs]{revtex4-1}

\usepackage{CJK}
\usepackage{amsmath}
\usepackage{amssymb}
\usepackage{graphics}
\usepackage{epsfig}
\usepackage{multirow}

\begin{document}

\title{Electronic thermoelectric power factor and Metal - Insulator transition in FeSb$_{2}$}
\author{Qing Jie,$^{1,\dag}$ Rongwei Hu,$^{1,\S}$ Emil Bozin,$^{1}$ A. Llobet,$^{2}$ I. Zaliznyak,$^{1}$ C. Petrovic$^{1,\ddag}$ and Q. Li$^{1,\ddag}$}

\affiliation{$^{1}$Condensed Matter Physics and Materials Science Department, Brookhaven National Laboratory, Upton, NY 11973 USA}
\affiliation{$^{2}$Lujan Neutron Scattering Center, LANL, MS H805, Los Alamos, New Mexico 87545, USA}

\date{\today}

\begin{abstract}
We show that synthesis-induced Metal -Insulator transition (MIT) for electronic transport along the orthorombic \textit{c} axis of FeSb$_{2}$ single crystals has greatly enhanced electrical conductivity while keeping the thermopower at a relatively high level. By this means, the thermoelectric power factor is enhanced to a new record high S$^{2}$$\sigma$ $\sim$ 8000 $\mu$WK$^{-2}$cm$^{-1}$ at 28 K. We find that the large thermopower in FeSb$_{2}$ can be rationalized within the correlated electron model with two bands having large quasiparaticle disparity, whereas MIT is induced by subtle structural differences. The results in this work testify that correlated electrons can produce extreme power factor values.
\end{abstract}


\maketitle

\section{Introduction}

The efficiency of thermoelectric (TE) material at temperature \textit{T} is evaluated by the figure of merit $ZT = (S^{2}/\rho\kappa)T$, where \textit{S} is the thermopower, $\rho$ is the electrical resistivity and $\kappa$ is the thermal conductivity. There are two distinct approaches to increasing the \textit{ZT}: either by $\kappa$ reduction or by power factor $(S^{2}/\rho$) enhancement. Techniques such as alloying, "phonon glass electron crystal" (PGEC) approach, and nanostructure engineering have been used to reduce the phonon mean free path, reduce lattice $\kappa$ and produce high \textit{Z}, at or above the room temperature.\cite{Ioffe,Sales,Nolas,Kim,Poudel} Reduction of the phonon mean free path has limitations since it cannot be reduced below the interatomic spacing and therefore other mechanisms for high \textit{ZT} are sought after.\cite{Cahill} Tuning the carrier density by doping is often inadequate since lower $\rho$ comes with higher carrier concentration that favors lower \textit{S}.\cite{Mahan1}

The most favorable TE electronic structure is the one that has a resonance in the density of states centered about 2-3 $k_{B}T$ away from the Fermi energy ($\epsilon$$_{F}$).\cite{Mahan2} Kondo Insulators (KI) represent a close approximation of such an ideal case. In a strongly correlated electron system, and, in particular, in a KI, localized \textit{f} or \textit{d} states hybridize with conduction electron states leading to the formation of a small hybridization gap. The density of states just below and just above the hybridization gap becomes very large. The thermopower is very sensitive to variations in density of states in the vicinity of the $\epsilon$$_{F}$, hence very large absolute values of $|S|$ $>$ 100 $\mu$V/K can be expected and are reported in KI.\cite{Takabatake} The materials include not only rare-earth based compounds but also FeSi.\cite{Sales2,Jones1,Jones2,Sato,Abe,Harutyunyan}

FeSb$_{2}$ crystallizes in \textit{Pnnm} orthorhombic structure and has been characterized as an example of strongly correlated non-cubic Kondo insulator-like material with \textit{3d} ions.\cite{Petrovic1,Petrovic2,Perucchi} Similar to FeSi, heavy fermion states were discovered in FeSb$_{2}$ by doping-induced metallization.\cite{BentienHF,DiTusa} Colossal values of thermopower up to $\sim$ 45 mV/K at 10 K and a record high thermoelectric power factor (TPF) of $\sim$ 2300 $\mu$W/K$^{2}$cm were observed.\cite{Bentien} This is two orders of magnitude larger than in best Bi$_{2}$Te$_{3}$-based materials and one order of magnitude larger than any reported value in correlated electron systems. Crystals in Refs. 19 and 21 exhibit semiconductor behavior and small resistivity anisotropy with relatively high resistivity for current directed along all three principal crystalline axes. In contrast, single crystals used in Kondo insulator studies showed high anisotropy of the resistivity: electric transport along the \textit{a} and \textit{b} axes is semiconducting, while the \textit{c} axis is metallic at high temperatures and exhibits metal-insulator transition (MIT) at 40 K.\cite{Petrovic1,Petrovic2,Hu1}

Here we investigated the TPF in two FeSb$_{2}$ crystals, with (crystal 1) and without (crystal 2) MIT. We show that large \textit{S} enhancement can be attributed to electronic correlations in multiple charge and heat carrying bands. We also provide evidence for structural origin of MIT in correlated electron bands that reduces $\rho$ by several orders of magnitude around 30 K in crystal 1 and results in a new record high TPF.

\begin{figure}
\centerline{\includegraphics[scale=0.7]{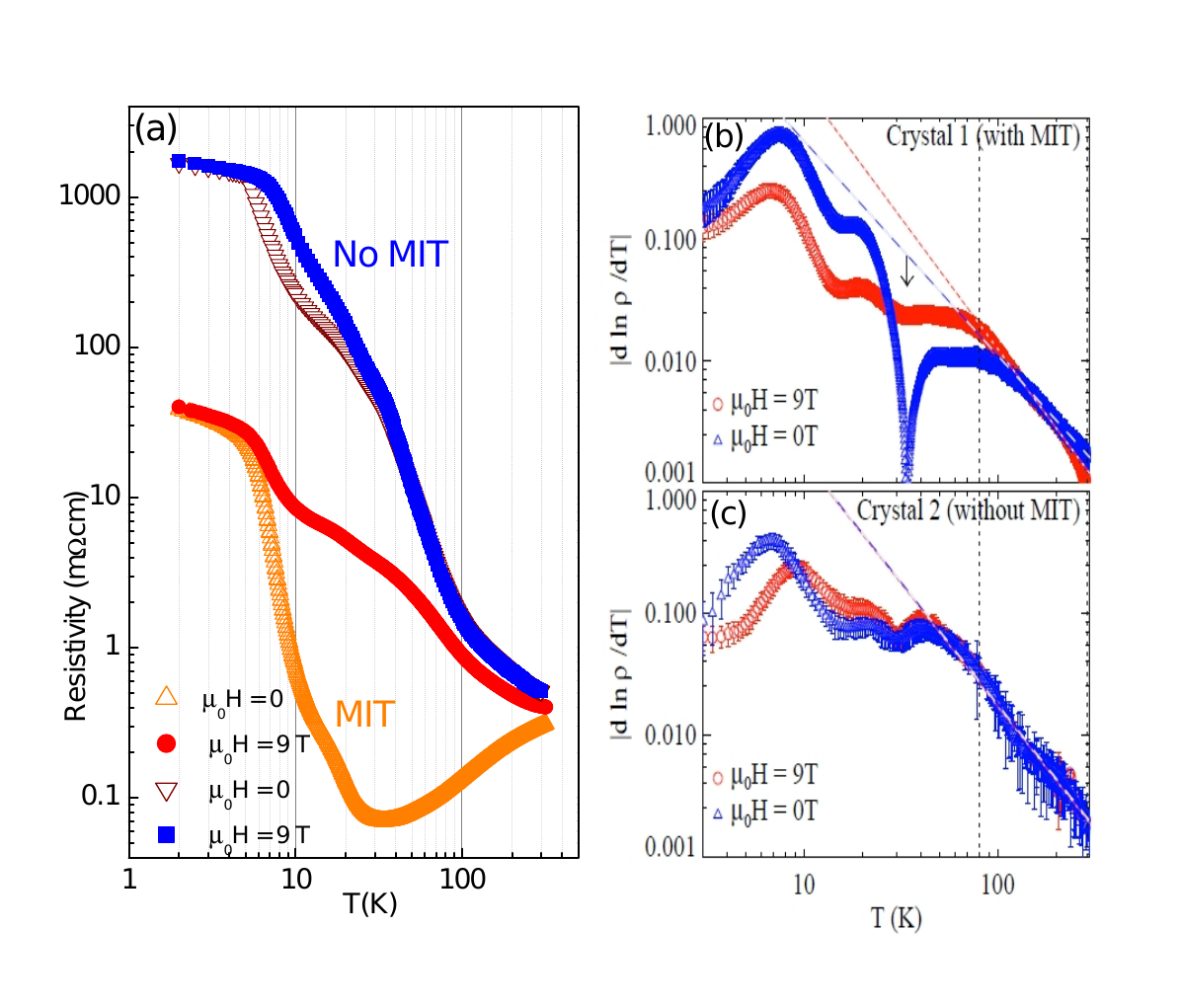}} \vspace*{-0.3cm}
\caption{(a) Comparison of \textit{c}-axis resistivity for two crystals. In 9 T  $\rho(T)$ in crystal 1 above 80 K is consistent with semiconducting thermally-activated transport with an activation gap of 120 K. (b),(c) Temperature regime(s) where $\rho(T)$ is dominated by thermally activated transport. The arrow in (b) marks the MIT in crystal 1, where $d[ln(\rho)]/dT$ changes sign. Broken lines in both panels are linear fits to activation-type behavior.}
\end{figure}

\section{Experiment}

Single crystals of FeSb$_{2}$ with MIT have been prepared as in Refs. 16 and 17 by decanting at 650$^\circ$C. Single crystals of FeSb$_{2}$ without MIT have been prepared by the method described in Refs. 19 and 21, i.e., by decanting at 690$^\circ$C, after a cool down from high temperatures to 640$^\circ$C. Crystals were oriented using Laue camera and cut into 0.6$\times$0.5$\times$4.5 (a$\times$b$\times$c) mm$^{3}$  samples for two probe thermopower measurements with two ends soldered on disk-shape leads using indium. Resistivity $\rho(T)$ was determined by a standard four-point ac method.  Heat and current transport along the orthorombic \textit{c} axis were measured using the Quantum Design PPMS platform. A magnetic field was applied along the crystallographic \textit{a} axis. Sample dimensions were measured with an optical microscope Nikon SMZ-800 with 10 $\mu$m resolution. Consequently the relative errors on the electrical resistivity and thermal conductivity are 4$\%$ and for the Hall coefficient 2$\%$.  Since the Seebeck coefficient does not depend on the sample geometry the main source of error is the sample uniformity and the accuracy of crystal orientation which introduces the measurement error of up to 5$\%$.

The atomic pair distribution (PDF) method, based on the total scattering approach, yields structural information on different length scales.\cite{Billinge} X-ray scattering experiments for PDF analysis were carried out at the 11-ID-C beamline of the Advanced Photon Source using high energy beam (\textit{E} = 114.82 keV, $\lambda$ = 0.108 ${\AA}$, 0.5 mm $\times$ 0.5 mm size). Both PDF, \textit{G(r)}, and its algebraic relative, radial distribution function (RDF), \textit{R(r)}, were considered in this study. Experimental PDF, \textit{G(r)} is obtained from the measured reduced total scattering structure factor, \textit{F(Q)=Q[S(Q)-1]}, via sine Fourier transform $G(r) = G(r)=(2/\pi)\int\limits_{0}^{\infty}(F(Q)Sin(Qr)dQ)$. In practice, the upper limit of integration is some finite value $Q_{max}$. RDF, \textit{R(r)} is obtained from \textit{G(r)} through: $R(r)$$=$$rG(r)+4$$\pi$$r^{2}$$\rho_{0}$ where $\rho_{0}$ is the average number density. Experimental setup for total scattering x-ray experiments utilized a Cryo Industries of America cryostat and Perkin-Elmer image plate detector. Finely ground samples in cylindrical polyimide capillaries were placed in a low-temperature sample changer, and the data for the two samples were successively collected for 4 min. at each temperature in probed range between 5 and 300 K. PDFs were obtained up to $Q_{max}$ = 26 {\AA}$^{-1}$ momentum transfer using standard protocols, and intermediate length scale structure modeled over (1.7 - 45.0) {\AA}$^{-1}$ range with \textit{Pnnm} structural model using the program PDFGUI.\cite{Farrow} Preliminary reference neutron total scattering based PDF's were obtained using a time-of-flight HIPD instrument at Los Alamos Neutron Scatttering Center.

\section{Results and Discussions}

The $\rho_{c}$ for crystal 1 is metallic down to the onset of MIT ($T_{MIT}$), as opposed to crystal 2 [Fig. 1(a)]. \cite{Petrovic1,Petrovic2,Bentien,Hu1} Above 80 K and up to 300 K, $\rho_{c}(T)$ of crystal 2 is insensitive to magnetic field and is semiconducting. Interestingly, the logarithmic derivative, $d[ln\rho_{c}(T)]/dT$ reveals an underdeveloped anomaly at $T_{MIT}$ in crystal 2 which points to the intrinsic nature of this temperature scale [Fig. 1(b,c)]. Arrhenius analysis assuming thermally activated behavior leads to a number of distinct energy scales valid for limited subranges of this temperature interval (Table I), allowing only approximate estimation of the gap values. The absolute value $|d[ln(\rho(T))]/dT|$ in the metallic and insulating states is very similar, which is reminiscent of the universality observed near the MIT in low-dimensional electron gas systems.\cite{Sodhi} Below 5 K electrical transport is dominated by extrinsic impurity states and is governed by a very small activation gap, similar for both crystals. Distinct energy scales and indirect gaps are in agreement with optical spectroscopy studies and \textit{ab initio} calculations. \cite{Perucchi,Bentien,Herzog,Lukoyanov} The large activation gap observed above 80 K in crystal 2 is not observed in the metallic phase of crystal 1, but its size argues against contribution of impurity states.

\begin{table}[b]
\caption{Approximate (see text) energy gap from electronic transport, the ratio of quasiparticle weights and scattering times in the valence and conduction band
$Z\Gamma = ([(Z{_v}^{2}\tau_{v}^{5/2})/\Gamma_{v}]/[(Z{_c}^{2}\tau_{c}^{5/2})/\Gamma_{c}$]) and $\epsilon$$_{F}$$\sim$ln($\mu$$_{c}$/$\mu$$_{v}$)
for a crystal with MIT (top) and a crystal with no MIT (bottom row).The huge difference of fitting parameters between
crystals 1 and 2 may be rationalized by the fact that the mobility difference of two bands for crystals with MIT is 10$^{8}$-10$^{9}$
while the one for crystals without MIT is $\sim$10, whereas both show similar large Seebeck coefficients.}
\begin{ruledtabular}
\begin{tabular}{ccccc}
(2-5)K  & (10 - 30)K & (80 - 100)K & $Z\Gamma$ & $\epsilon$$_{F}$\\
0.14(1) meV & 6.2(1) meV & Metallic & 6.2$\cdot$10$^{-8}$ & 3.4 meV \\
0.08(1) meV & 4.1(1) meV & 14.7(1) meV & 38 & 0.8 meV
\end{tabular}
\end{ruledtabular}
\end{table}

\begin{figure}
\centerline{\includegraphics[scale=0.7]{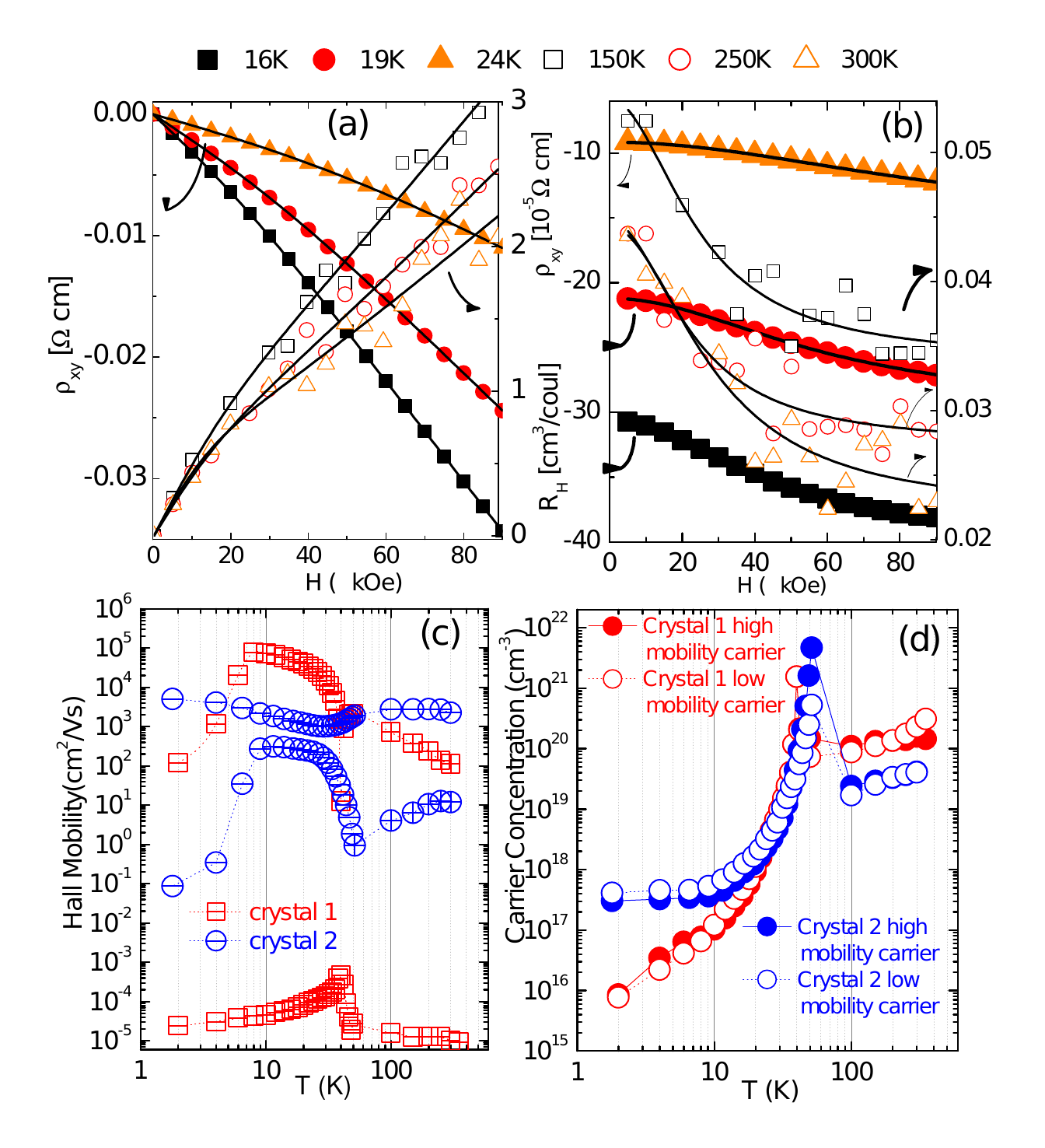}} \vspace*{-0.3cm}
\caption{The $\rho$$_{xy}$ (a) and Hall constant (b) for crystal 2. Hall mobility (c) where the hole (electron) carriers are denoted by + (-) and carrier concentrations of high and low mobility carriers (d) as a function of temperature for both crystals.}
\end{figure}

The Hall resistivity $\rho_{xy}(B)$ (current flowing along the \textit{c} axis) for both crystals [Fig 2 (a,b)] is nonlinear, confirming the presence of two carrier bands. In the absence of skew scattering we proceed to the analysis of the data in the two-band picture. In a two-carrier system \cite{KimJS} the Hall coefficient is $R_{H} = \rho_{xy}/H = \rho_{0}(\alpha_{2}+\beta_{2}H^{2})/(1+\beta_{3}H^{2})$ where $\alpha_{2} = f_{1}\mu_{1} + f_{2}\mu_{2}$, $\beta_{2} = (f_{1}\mu_{2} + f_{2}\mu_{1})\mu_{1}\mu_{2}$ and $\beta_{3} = (f_{1}\mu_{2} + f_{2}\mu_{1})^{2}$ where $\rho_{0}$ = $\rho(B=0)$, $f_{i} = |n_{i}\mu_{i}|/\Sigma|n_{i}\mu_{i}|$ is the f factor, $n_i{}$ and $\mu_{i}$ are individual carrier band concentrations and  mobilities. The agreement with the model is excellent and we obtain carrier band concentrations and mobilities for crystal 2 while also showing data for crystal 1 for comparison.\cite{Hu2} In the model "carrier" denotes a set of carriers with identical mobility associated with only one energy and/or one degenerate energy level and is different from the conventional electron or hole carriers that correspond to a continuous energy band. The low mobility carriers in crystal 2 are hole type ($\mu(2)_{L}\sim$ 10 cm$^{2}$/Vs) and they are an order of magnitude less mobile when compared to hole carriers in crystal 1 at 300 K. Note that hole carriers in crystal 1 at room temperature constitute high mobility band ($\mu(1)_{H}\sim$ 10$^{2}$ cm$^{2}$/Vs). Yet, just like $\mu(1)_{H}$, the $\mu(2)_{L}$ band exhibits sign change below 50 K, having identical temperature dependence when compared to $\mu(1)_{H}$ with up to three orders of magnitude lower nominal values. The $\mu(2)_{H}$ carriers are electron type in crystal 2. They show little change in magnitude and do not change sign. The most striking difference between crystals 1 and 2 is the ratio of nominal mobility values. The large mobility difference is absent in crystal 2. Moreover, there is about or less than one order of magnitude between $\mu(2)_{H}$  and $\mu(2)_{L}$ in the high thermopower region from 10 to 40 K. The low and high carrier concentrations in both crystals have nearly identical values, suggesting a compensated nature of electronic transport at all temperatures. The carrier concentrations $n_{2}(H)$ and $n_{2}(L)$ are rather close at 300 K but they are both about an order of magnitude lower and have the same temperature dependence as carrier concentrations $n_{1}(H)$ and $n_{1}(L)$ in crystal 1. Carrier concentrations in crystals 1 and 2 are nearly identical in the temperature region of high thermopower, whereas in the impurity regime below 10 K crystal 2 (no MIT) shows much higher carrier concentrations. Values of carrier mobilities for crystal 2 ($\sim$10$^{-3}$ cm$^{2}$/Vs) are in general agreement with mobility values at low temperatures obtained using a two-band model (4 - 25 K) on crystals with no MIT.\cite{Takahashi2}

Using carrier concentrations and Hall mobilities from a large number of measured $\rho_{xy}(B)$ isotherms we obtain \textit{S(T)} for both crystals in the framework of a two-band non-interacting semiconductor \cite{Saunders,Putley,Jovovic} model where thermopower
$S=(S_{e}\sigma_{e}+S_{h}\sigma_{h})/(\sigma_{e}+\sigma_{h}$) and
\begin{equation}
S_{e,h}=(k_{B}/e)[\frac{((5/2)+s)F_{(3/2)+s}(\xi_{e,h})}{((3/2)+s)F_{(1/2)+s}(\xi_{e,h})}]-\xi_{e,h}
\end{equation}
where $F_{j}(\xi)=\int\limits_{0}^{\infty}\frac{x^{j}dx}{1+exp(x-\xi)}$. The scattering exponent $s$ represents the energy dependence of the relaxation time $\tau$=$\tau_{0}$$\epsilon^{s}$, $\xi$=$\epsilon_{F}$/(k$_{B}$T) are the reduced Fermi energies for electrons and holes $\xi_{e}$ and $\xi_{h}$ (as measured from the bottom of the conduction band for electrons and from the top of the valence band for holes) and $\epsilon_{F}$ = h$^{2}$/(2m$^{*}$)(N/V)$^{(2/3)}$(3/8$\pi$)$^{(2/3)}$. Hence, for semimetals $\epsilon_{Fe}$ = - ($\epsilon_{Fh}$ + $\epsilon_{0})$ and $\xi_{e}$= - $\xi_{h}$ - $\epsilon_{0}$/(k$_{B}$T) where $\epsilon_{0}$ is the overlap energy, and for semiconductors $\epsilon_{Fe}$ = - $\epsilon_{Fh}$ + $\epsilon_{g}$ and $\xi_{e}$ = - $\xi_{h}$ - $\epsilon_{0}$/(k$_{B}T)$ where $\epsilon_{g}$ is the energy gap for semiconductors. In this model we used value \textit{s = -1/2} assuming that acoustic phonon scattering is dominant at all temperatures and $m^{*}$ = $m_{e}$. The two-band non-interacting semiconductor model adequately describes crystal 2 above 50 K and below 5 K, whereas it fails to explain $|S(T)|$ for crystal 1 in its metallic state (Fig. 3). While the similar shape of calculated and measured thermopower for both crystals between 10 and 40 K in the noninteracting model argues in favor of the two-band approach, such calculation does not explain the amplification of $|S(T)|$ observed in the (10-30) K range. The prime suspect for discrepancy are strong correlations which can have a significant impact on carriers in \textit{3d} bands.

In a correlated electronic system with two bands separated by a gap, \textit{S} depends not only on the gap size but also on the anisotropy or asymmetry in the transport function:\cite{Tomczak}
\begin{equation}
S=\frac{1}{|e|T}(\epsilon_{F}-\frac{\Delta}{2}\delta\lambda)-\frac{5k_{B}}{2|e|}\delta\lambda
\end{equation}
where $\epsilon$$_{F}$ is the chemical potential, $\Delta$ is the gap and $\delta$$\lambda$=($\lambda$$_{c}$-$\lambda$$_{v}$)/($\lambda$$_{c}$+$\lambda$$_{v}$) is the asymetry parameter. The asymmetry parameter carries the information on band specific correlation strengths since $\lambda$$_{c,v}$=$Z_{c,v}^{2}$m$_{c,v}^{*(5/2)}$e$^{-\beta\mu}$/$\Gamma$$_{c,v}$m$_{0}^{2}$ where $Z_{c,v}$ are quasiparticle weights, $\Gamma$$_{c,v}$ are scattering amplitudes and $m_{c,v}^{*}$ is the effective mass of carriers in conduction and valence bands. In addition, in real materials the $|S(T)|_{max}$ rapidly diminishes with the increase in impurity concentration due to the increase of scattering amplitude, as recently observed in FeSb$_{2}$.\cite{Tomczak,Bentien2,Takahashi} Bandwidth narrowings $m_{c,v}^{*}$ influence \textit{S(T)} not only through the asymmetry parameter $\delta\lambda$ but also via chemical potential since $\epsilon_{F}$=(3k$_{B}$T/4)ln(m$_{v}^{*}/m_{c}^{*})$.\cite{Tomczak} Finally, the presence of impurities can have considerable effect on both $|S(T)|_{max}$ and \textit{S(T)}. The \textit{S(T)} magnitude at a given temperature depends on the ability of impurity carriers to put the Fermi level in the optimal position for thermopower enhancement.\cite{Tomczak}

Both crystals have similar \textit{S(T)} and nearly identical magnetothermopower \textit{MT = [S(9T)-S(0)]/S(0)]} in the region of high \textit{S} (Fig. 3). In contrast to differences in $\rho(T)$, thermopower \textit{S(T)} for heat flow along the \textit{c}-axis is rather similar above 120 K and below 8 K. Thermopower changes sign from positive to negative above 100 K in both crystals, indicating the presence of two carrier types. Since $m_{c,v}^{*}$=e$\tau_{c,v}$/$\mu_{c,v}$  where $\tau_{c,v}$ is the scattering time and $\mu_{c,v}$ is the mobility in conduction and valence bands, we use mobility values for individual bands obtained from the fits of the $R_{H}$ in the two-band model. The value of $Z^{2}$$\tau$$^{5/2}$/$\Gamma$ then becomes a fit parameter for each carrier band, in addition to the chemical potential $\epsilon$$_{F}$.  Fits to the above equation for a fixed value of the gap which corresponds to the correlated electron temperature region (10-30) K are shown in Fig. 3(a) as red (crystal 1) and blue (crystal  2) solid lines. The best fits are obtained for the gap values of $\sim$ (15-20) meV (Table I), suggesting that the large enhancement of \textit{S} is due to the strong electronic correlations associated with a smaller indirect gap.\cite{Perucchi,Herzog} Note that this is within the extremal limits in the assymetric case when the chemical potential is near the edge of the valence band ($|S(T)e|$$\leq$ $\Delta$/T+5/2k$_{B}$).\cite{Tomczak} For large $\Gamma$ or for similar $\lambda$$_{c}$ and $\lambda$$_{v}$ (symmetric multiband effects) $S\rightarrow$0. This is observed: \textit{S} sign change is at $T = (120 \pm$ 4) K, for both crystals 1 and 2.  This implies that correlation effect on \textit{S(T)} vane at high temperatures.

The phonon drag is unlikely to have a significant contribution to \textit{S} since isostructural RuSb$_{2}$ and FeAs$_{2}$ have larger $\kappa$ and much smaller values of $|S(T)|$$_{max}$ due to different temperature dependence of \textit{S} and Nernst coefficient.\cite{Perucchi,Lazarevic,Sun1,Sun2} The $|$S(T)$|$ rises below 40 K, reaching values of 1 mV/K (crystal 1) and 1.9 mV/K (crystal 2) in the (10-20) K range (Fig. 4). The $S^{2}$$\sigma$ of crystal 1 (7800 $\mu$WK$^{-2}$cm$^{-1}$) is maximized at $T_{max}$ = 28 K, which is higher than in crystal 2 ($T_{max}$ = 20 K) and close to T$_{MIT}$. The maximum value we find is three times larger than the TPF previously observed\cite{Bentien} in FeSb$_{2}$ and occurs at 16 K higher temperature.

\begin{figure}
\centerline{\includegraphics[scale=0.6]{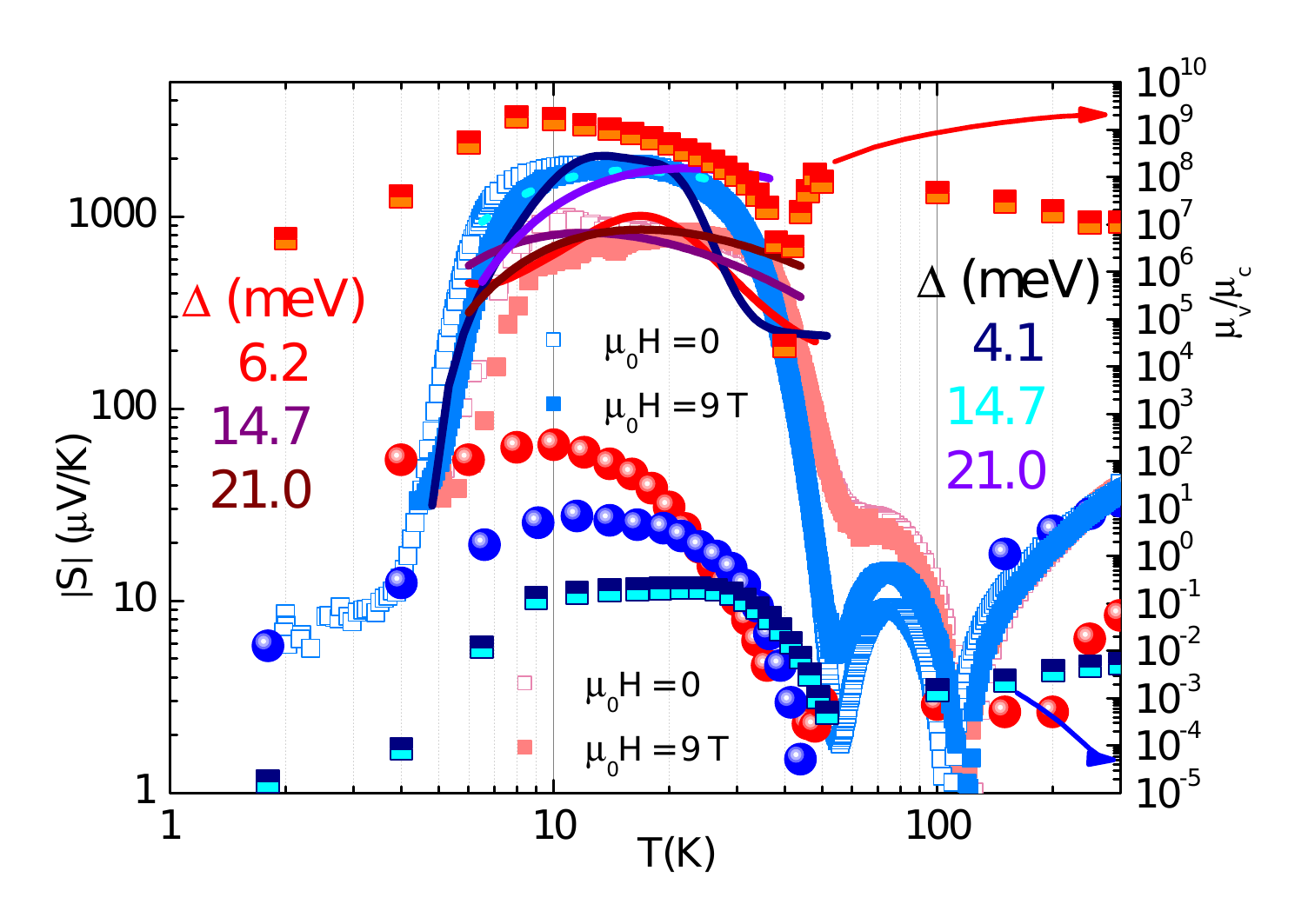}} \vspace*{-0.3cm}
\caption{(a) $|S(T)|$ for FeSb$_{2}$ crystals 1 (red) and 2 (blue squares). Red and blue balls show fits to the two band noninteracting semiconductor model. Fits to the correlated electron model for crystal 1 are shown by red, purple and brown (crystal 1) and dark blue, light blue and violet (crystal 2) solid lines that correspond to increasing values of fixed energy gap $\Delta$ (see text). The ratio of individual band mobilities in a two-carrier model is shown by the red/orange (crystal 1) and dark/light blue squares (crystal 2).}
\end{figure}

\begin{figure}
\centerline{\includegraphics[scale=0.65]{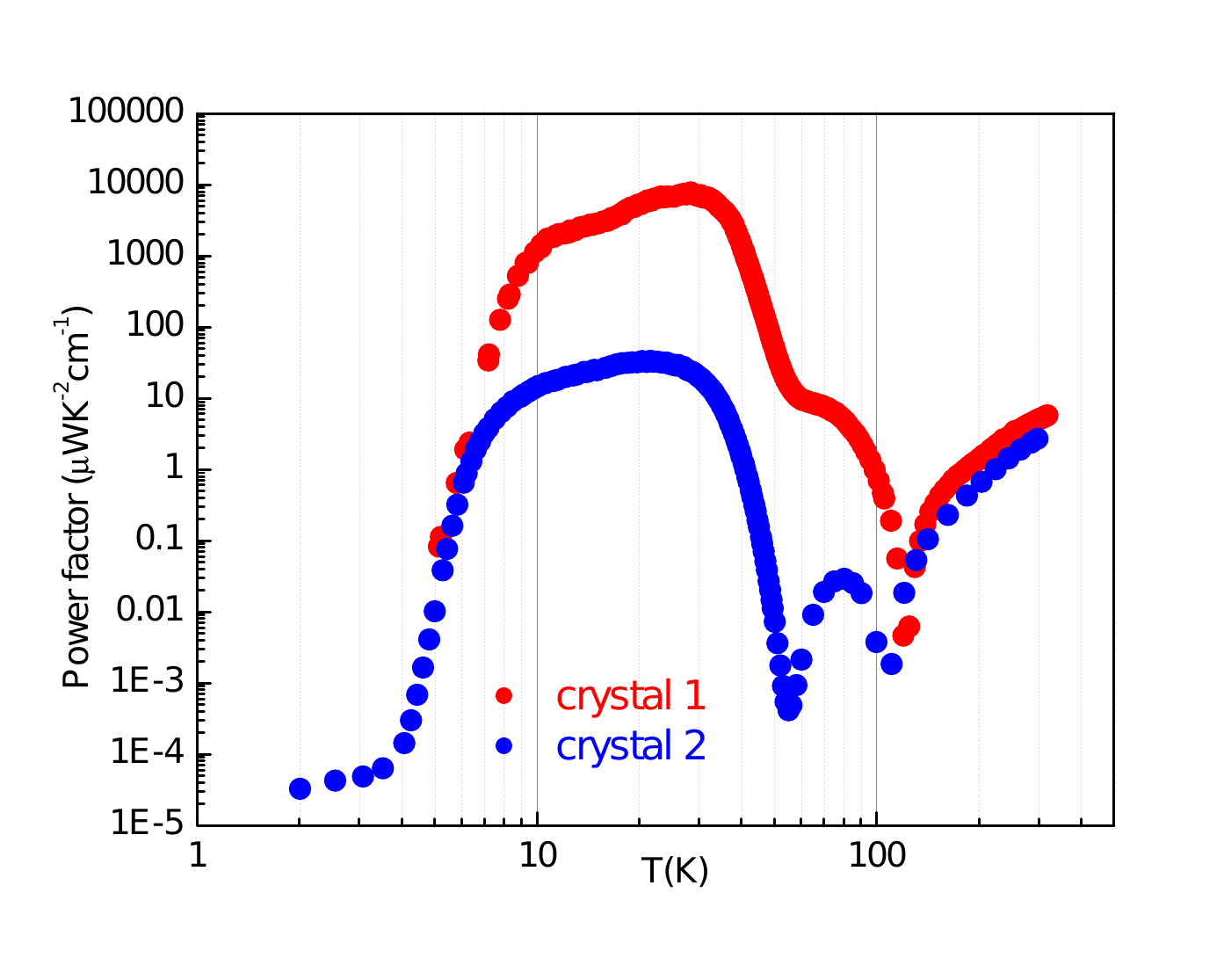}} \vspace*{-0.3cm}
\caption{(The low resistivity around MIT leads to a record high TPF. Crystal 1 has two orders of magnitude higher TPF between 8 and 100 K.}
\end{figure}

The central finding of the above analysis is that large thermopower in both FeSb$_{2}$ crystals can be rationalized within the correlated electron model with two bands having the large disparity of quasiparticle properties. This is best illustrated by the ratio of carrier mobilities in the conduction and valence bands. While the mobility of the valence band carriers in crystal 1 greatly exceeds that of the conduction carriers, they are practically immobilized in crystal 2 where conduction carriers have higher mobility. A key observation here is that despite great disparity in absolute values, mobilities of the valence band carriers in two crystals show strikingly similar temperature dependence with a pronunced anomaly at MIT. Most importantly, in both cases there is a change in the nature of valence charge carriers, which are holes at temperatures above MIT and electrons at lower temperatures. As we discuss later, this observation reveals the likely nature of the MIT and of its absence in crystal 2.

Now we turn to the quasi-one-dimensional (quasi-1D) metallic conductivity\cite{Petrovic1} and the MIT in crystal 1 which is the key for colossal TPF. For a large concentration of impurities the chemical potential can go into the valence/conduction band, producing metallic $\rho(T)$.\cite{Tomczak} While an order of magnitude disparity in the number of charge carriers observed in crystals 1 and 2 above $\sim$100 K would be consistent with such a scenario, it cannot be reconciled with the closely compensated nature of the carrier content. Below 10 K, where conductivity is governed by impurities, there are more carriers in crystal 2 than in crystal 1. Finally, impurity bands cannot account for the quasi-1D metallic conductance. Band structure suggests that the likely origin of the quasi-1D transport is the non-bonding $d_{xy}$ band, where the overlaps of Fe $d_{xy}$ orbitals are along the chains of edge-sharing octahedra parallel to the \textit{c} axis, with little or no overlaps between orbitals in different chains [Fig. 5 (a)].\cite{Lukoyanov,Goodenough,Hulliger} Hall data shows that quasi-1D metallic conductance in sample 1 at temperatures above 40 K is provided by a small number of mobile holes, which implies a nearly filled valence band. Such a non-bonding band must be narrow and strongly correlated, and should be described by the 1D Hubbard model.\cite{Essler} The low-dimensional transport in such band is sensitive to disorder and MIT is expected at half filling. Can it be the origin of the MIT observed in crystal 1 that $d_{xy}$ band is depleted with the decreasing temperature, attaining half-filling at $T_{MIT}$? Could it then be that a small additional disorder in crystal 2 induces localization of the quasi-1D charge carriers?

\begin{figure}
\centerline{\includegraphics[scale=0.75]{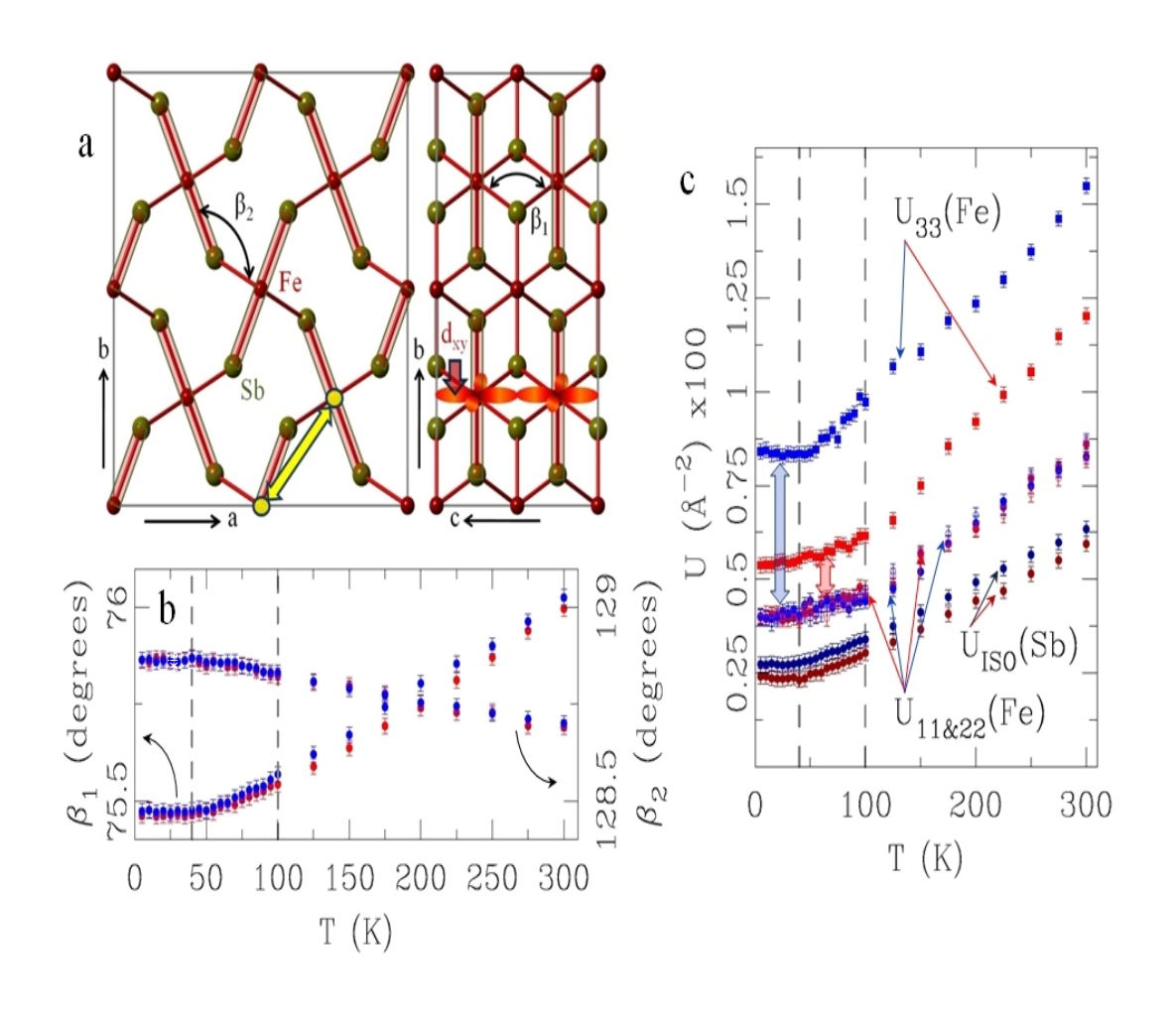}} \vspace*{-0.3cm}
\caption{(Structural unit of FeSb$_{2}$ (a) shows 2x2x2 \textit{Pnnm} unit cell as seen from the top of the \textit{c}-axis (left), and from the top of the \textit{a}-axis (right). Shaded rectangles denote short Fe-Sb distances within the FeSb$_{6}$ octahedra. Intrachain and interchain Fe-Sb-Fe angles are denoted as $\beta$$_{1}$ and $\beta$$_{2}$, respectively. Next-nearest-neighbour interchain Fe-Fe distance is indicated by the double arrow. (b) Temperature dependence of $\beta_{1}$ and $\beta_{2}$ for crystal 1 (red) and crystal 2 (blue). (c) ADP factors for Fe (anisotropic) and Sb (isotropic) for crystal 1 (red) and  crystal 2 (blue). Vertical dashed lines in (b) and (c) at 40K and 100K indicate temperatures of insulator-metal transition and change of the nature of the charge carriers, respectively.}
\end{figure}

In order to answer this question, we performed the PDF analysis of the x-ray data taken on powder samples of crystals 1 and 2 (Figs. 5 and 6). The PDF provides insight into the crystal structure on short and intermediate length-scales and allows quantifying subtle structural features such as bond properties and local disorder. Intrachain and interchain Fe-Sb-Fe angles, $\beta$$_{1}$(T) and $\beta$$_{2}$(T) respectively, shown in Fig. 5(b), are very similar, with a small difference developing in $\beta$$_{1}$ above 100 K. This may indicate charge redistribution involving $d_{xy}$ bands since the presence of Fe $d_{xy}$ charge will result in a somewhat larger $\beta_{1}$ angle.\cite{Hulliger} Atomic displacement factors (ADP) [Fig. 5(c)] of Fe atoms are rather anisotropic: mean-square atomic displacements along the \textit{a} and the \textit{b} axes, $U_{11}$ and $U_{22}$, are nearly equal for both crystals, while those along the \textit{c} axis, $U_{33}$, are noticeably enhanced in crystal 2 at all temperatures, being 0.0084 ${\AA}^{2}$ (crystal 2) and ~0.0054 ${\AA}^{2}$ (crystal 1) at 10 K. Enhanced ADPs are typical indication of disorder. In this case disorder is markedly 1D, precisely along the \textit{c} axis and significantly higher in crystal 2.

It is clear from the difference curves (Fig. 6) in the x-ray RDF (Refs. 23 and 24) that the average local environments on the length scale of about one \textit{Pnnm} unit cell (reflected in the first several RDF peaks up to 4.5 ${\AA}$) are indistinguishable. On larger length scales structural features start to differ. The RDF peak at about 4.65 ${\AA}$, denoted by a vertical arrow in Fig. 6, is markedly broader and less intense for crystal 2 at all temperatures studied. This peak corresponds to the next-nearest-neighbor Fe-Fe distance [FeSb$_{6}$ interchain distance, double arrow in Figure 5(a)], with no other contributions. Such a discrepancy could have several origins arising in a difference in Fe-Fe bondlength distribution, amount of Fe in the structures of the two crystals, charge state of Fe between the two crystals, or a combination of these effects.

\begin{figure}
\centerline{\includegraphics[scale=0.7]{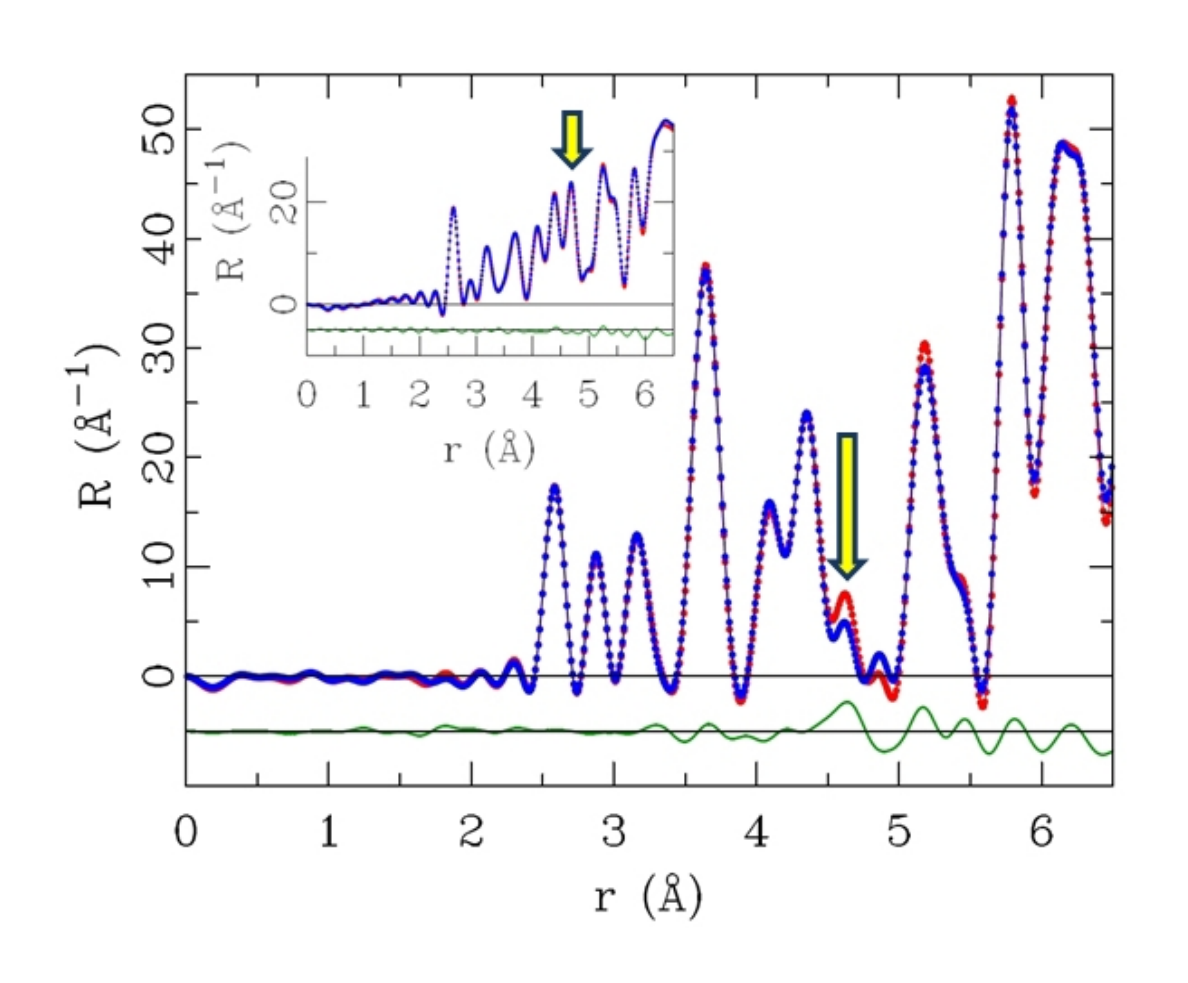}} \vspace*{-0.3cm}
\caption{Experimental neutron RDF data (inset) at 20 K reveals no detectable difference, suggesting that the disorder observed in x-ray experiment probably originates from the charge sector.}
\end{figure}

Low-dimensional $\rho$ is very sensitive to disorder which, in the simplest picture, suppresses metallic state by inducing strong localization. More generally, the disorder can impact the $d_{xy}$ overlap and the band structure, the orbital character of the electronic states responsible for conduction, the occupancy of Fe $d_{xy}$ orbitals, and even the orbital-dependent Hubbard \textit{U} interaction strength in a $d_{xy}$ quasi-1D band of itinerant states that forms with sufficient $d_{xy}$ overlap. As opposed to this "self-generated" impurity level arising from Fe \textit{d} orbitals in crystals with MIT, extrinsic impurity levels cannot produce metallic resistivity near or close to room temperature as in crystals with MIT and are significant only at low temperatures.\cite{Hu2,Takahashi2}

Crystals with MIT were cooled slowly from 1000 $^\circ$C to 650 $^\circ$C and were decanted at that temperature
from the liquid Sb flux. On the other hand, crystals with no MIT were cooled to 640$^\circ$C, closer to Sb
solidification temperature and were subsequently decanted at 690$^\circ$C. There are two main factors that could contribute to increased
crystallographic disorder in crystals with no MIT. FeSb$_{2}$ melts and decomposes at 738$^\circ$C to FeSb and Sb. Hence, crystallization occurs over a rather narrow temperature window. Crystals decanted at 690$^\circ$C are much closer to the melting point when compared to crystals decanted at 650$^\circ$C. This could contribute to increased disorder in Fe-Sb chemical bonds. In addition, crystals cooled to 640 $^\circ$C are closer to solidification line and are likely to experience more stress from the flux. This is consistent with observed structural differences, suggesting that MIT in FeSb$_{2}$ crystals is governed by subtle structural differences tunable by synthesis procedure. Greatly increased conductivity near MIT, combined with significant electronic TEP, leads to a new record high thermoelectric power factor.

\section{Conclusion}

In summary, we report the highest known TPF induced by the synthesis-controlled MIT in the correlated electron semiconductor FeSb$_{2}$. The large thermopower enhancement can be understood within the electronic model, whereas MIT likely originates in quasi-1D and strongly correlated narrow band of itinerant states very sensitive to disorder. This is further supported by the recent observation of quasi-1D magnetism in isostructural CrSb$_{2}$.\cite{Stone}

\section{Acknowledgements}

We thank T. M. Rice and Simon Billinge for useful discussion and Milinda Abeykoon and Pavol Juhas for help with x-ray experiments. This work was carried out at the Brookhaven National Laboratory, which is operated for the U.S. Department of Energy by Brookhaven Science Associates DE-Ac02-98CH10886. The Advanced Photon Source at Argonne National Laboratory, operated by UChicago Argonne LLC, is supported under the U.S. DOE-OS Contract No. DE-AC02-06CH11357. Neutron PDF experiments were carried out on HIPD at LANSCE, funded by DOE BES; LANL is operated by Los Alamos National Security LLC under DE-AC52-06NA25396.

\dag Present address: Department of Physics, Boston College, Chestnut Hill, Massachusetts 02467
\S Present address: Department of Physics, University of Maryland, College Park,
MD 20742-4111

\ddag\ petrovic@bnl.gov and qiangli@bnl.gov


\begin{thebibliography}{99}

\bibitem{Ioffe} A. F. Ioffe, \textit{Physics of Semiconductors} (Academic, New York, 1960), p. 282

\bibitem{Sales} B. C. Sales, D. Mandrus, B. C. Chakoumakos, V. Keppens, and J. R. Thompson, Phys. Rev. B \textbf{56}, 15081 (1997).

\bibitem{Nolas} G. S. Nolas, J. L. Cohn, G. A. Slack, and S. B. Schujman, Appl. Phys. Lett. \textbf{73}, 178 (1998).

\bibitem{Kim} W. Kim, J. Zide, A. Gossard, D. Klenov, S. Stemmer, A. Shakouri, and A. Majumdar, Phys. Rev. Lett. \textbf{96}, 045901 (2006).

\bibitem{Poudel} B. Poudel, Q. Hao, Y. Ma, Y. Lan, A. Minnich, B. Yu, X. Yan, D. Wang, A. Muto, D. Vashaee, X. Chen, J. Liu, Mildred S. Dresselhaus, G. Chen, and Zhifeng Ren, Science \textbf{320}, 634 (2008).

\bibitem{Cahill} D. G. Cahill, S. K. Watson and R. O. Pohl, Phys. Rev. B \textbf{46}, 6131 (1992).

\bibitem{Mahan1} G. D. Mahan, B. Sales and J. Sharp, Physics Today, \textbf{50}, 42 (1997).

\bibitem{Mahan2} G. D. Mahan and J. O. Sofo, Proc. Natl. Acad. Sci. USA  \textbf{93}, 7436 (1996).

\bibitem{Takabatake} T. Takabatake, T. Sasakawa, J. Kitagawa, T. Suemitsu, Y. Echizen,
K. Umeo, M. Sera and Y. Bando, Physica B \textbf{328},53 (2003).

\bibitem{Sales2} B. C. Sales, E. C. Jones, B. C. Chakoumakos, J. A. Fernandez-Baca, H. E. Harmon, J. W. Sharp and E. H. Volckmann, Phys. Rev. B \textbf{50}, 8207 (1994).

\bibitem{Jones1} C. D. W. Jones, K. A. Regan, and F. J. DiSalvo, Phys. Rev. B \textbf{58}, 16057 (1998).

\bibitem{Jones2} C. D. W. Jones, K. A. Regan, and F. J. DiSalvo, Phys. Rev. B \textbf{60}, 5282 (1999).

\bibitem{Sato} H. Sato, Y. Abe, H. Okada, T. D. Matsuda, K. Abe, H. Sugawara, and Y. Aoki, Phys. Rev. B \textbf{62}, 15125 (2000).

\bibitem{Abe} K Abe, H Sato, T D Matsuda, T Namiki, H Sugawara and Y Aoki, J. Phys.: Condens. Matter \textbf{14}, 11757 (2002).

\bibitem{Harutyunyan} S. R. Harutyunyan, V. H. Vardanyan, A. S. Kuzanyan, V. R. Nikoghosyan, S. Kunii, K. S. Wood, and A. M. Gulian, Appl. Phys. Lett. \textbf{83}, 2142 (2003).

\bibitem{Petrovic1} C. Petrovic, J. W. Kim, S. L. Bud'ko, A. I. Goldman, P. C. Canfield, W. Choe and G. J. Miller,, Phys. Rev. B \textbf{67}, 155205 (2003).

\bibitem{Petrovic2} C. Petrovic, Y. Lee, T. Vogt, N. Dj. Lazarov, S. L. Bud$'$ko, and P. C. Canfield,  Phys. Rev. B \textbf{72}, 045103 (2005).

\bibitem{Perucchi} A. Perucchi, L. Degiorgi, Rongwei Hu, C. Petrovic, and V.F. Mitrovic, Eur. Phys. J. B \textbf{54}, 175 (2006).

\bibitem{BentienHF} A. Bentien, G. K. H. Madsen, S. Johnsen, and B. B. Iversen, Phys. Rev. B \textbf{74}, 205105 (2006).

\bibitem{DiTusa} J. F. DiTusa, K. Friemelt, E. Bucher, G. Aeppli and A. P. Ramirez, Phys. Rev. B \textbf{58}, 10288 (1998).

\bibitem{Bentien} A. Bentien, S. Johnsen, G. K. H. Madsen, B. B. Iversen and F. Steglich, Europhys. Lett. \textbf{80}, 17008 (2007).

\bibitem{Hu1} Rongwei Hu, K. J. Thomas, Y. Lee, T. Vogt, E. S. Choi, V. F. Mitrovic, R. P. Hermann, F. Grandjean, P. C. Canfield, J. W. Kim, A. I. Goldman, and C. Petrovic, Phys. Rev. B \textbf{77}, 085212 (2008).

\bibitem{Billinge} S. J. L. Billinge and T. Egami, \textit{Underneath the Bragg peaks: structural analysis of complex materials}, Pergamon Press Elsevier, Oxford England, (2003).

\bibitem{Farrow} C L Farrow, P Juhas, J W Liu, D Bryndin, E S Bozin, J Bloch, T. Proffen and S J L Billinge, J. Phys: Condens. Mat. \textbf{19}, 335219 (2007).

\bibitem{Sodhi} S. L. Sodhi, S. M. Girvin, J. P. Carini and D. Shahar, Rev. Mod. Phys. \textbf{69}, 315 (1997).

\bibitem{Herzog} A. Herzog, M. Marutzky, J. Sichelschmidt, F. Steglich, S. Kimura, S. Johnsen and B. B. Iversen, Phys. Rev. B \textbf{82}, 245205 (2010).

\bibitem{Lukoyanov} A.V. Lukoyanov, V.V. Mazurenko, V.I. Anisimov, M. Sigrist, and T.M. Rice, Eur. Phys. J. B \textbf{53}, 205 (2006).

\bibitem{KimJS} J. S. Kim, J. Appl. Phys. \textbf{86}, 3187 (1999).

\bibitem{Hu2} Rongwei Hu, V. F. Mitrovic and C. Petrovic, Appl. Phys. Lett. \textbf{92}, 182108 (2008).

\bibitem{Takahashi2} H. Takahashi, R. Okazaki, Y. Yasui, and I. Terasaki, Phys. Rev. B \textbf{84}, 205215 (2011).

\bibitem{Saunders} G. A. Saunders and O. Oktu, J. Phys. Chem. Solids, \textbf{29}, 327 (1968).

\bibitem{Putley} E. H. Putley, \textit{The Hall effect in Semiconductor Physics} (Dover, New York 1968).

\bibitem{Jovovic} V. Jovovic and J. P. Heremans, Phys. Rev. B \textbf{77}, 245204 (2008).

\bibitem{Tomczak} Jan M. Tomczak, K. Haule, T. Miyake, A. Georges, and G. Kotliar, Phys. Rev. B \textbf{82}, 085104 (2010).

\bibitem{Bentien2} A. Bentien, G. K. H. Madsen, S. Johnsen, and B. B. Iversen, Phys. Rev. B \textbf{74}, 205105 (2006).

\bibitem{Takahashi} H. Takahashi, Y. Yasui, I. Terasaki and M. Sato, J. Phys. Soc. Japan \textbf{80}, 054708 (2011)

\bibitem{Lazarevic} N. Lazarevic, Z. V. Popovic, Rongwei Hu and C. Petrovic, Phys. Rev. B \textbf{81}, 144302 (2010).

\bibitem{Sun1} P. Sun, N. Oeschler, S. Johnsen, B. B. Iversen, and Frank Steglich, Appl. Phys. Express \textbf{2}, 091102 (2009).

\bibitem{Sun2} Peijie Sun, Niels Oeschler, Simon Johnsen, Bo Brummerstedt Iversen, and Frank Steglich, Phys. Rev. B \textbf{79}, 153308 (2009).

\bibitem{Goodenough} J. B. Goodenough, J. Solid State Chem. \textbf{5}, 144 (1972).

\bibitem{Hulliger} F. Hulliger Struct. Bonding (Berlin) \textbf{4}, 83 (1967).

\bibitem{Essler} F. H. L. Essler, \textit{The One Dimensional Hubbard Model}, Cambridge University Press (2005).

\bibitem{Stone} M. B. Stone, M. D. Lumsden, S. E. Nagler, D. J. Singh, J. He, B. C. Sales, and D. Mandrus, Phys. Rev. Lett. \textbf{108}, 167202 (2012).

\end{thebibliography}
\end{document}